\documentclass[prb, twocolumn]{revtex4}

\usepackage{graphicx}
\usepackage{xcolor}
\usepackage{amsmath}
\usepackage{amssymb}
\usepackage{siunitx}
\usepackage[english]{babel}
\begin{document}
\title{Single, double, and triple quantum dots in Ge/Si nanowires}
\author{F. N. M. Froning$^{1}$, M. K. Rehmann$^{1}$, J. Ridderbos$^{2}$, M. Brauns$^{2}$, F. A. Zwanenburg$^{2}$, A. Li$^{3}$, E. P. A. M. Bakkers$^{3,4}$, D. M. Zumb\" uhl$^{1}$, and F. R. Braakman$^{1\ast}$}

\affiliation{\vspace{1em} 1: Department of Physics, University of Basel, Klingelbergstrasse 82, 4056 Basel, Switzerland}
\affiliation{2: NanoElectronics Group, MESA+ Institute for Nanotechnology, University of Twente, P.O. Box 217,7500 AE Enschede, The Netherlands}
\affiliation{3: Department of Applied Physics, Eindhoven University of Technology, P.O. Box 513, 5600 MB Eindhoven, The Netherlands}
\affiliation{4: QuTech and Kavli Institute of Nanoscience, Delft University of Technology, 2600 GA Delft, The Netherlands}
\date{\today}

\begin{abstract}\noindent We report highly tunable control of holes in Ge/Si core/shell nanowires (NWs). We demonstrate the ability to create single quantum dots (QDs) of various sizes, with low hole occupation numbers and clearly observable excited states. For the smallest dot size we observe indications of single-hole occupation. Moreover, we create double and triple tunnel-coupled quantum dot arrays. In the double quantum dot configuration we observe Pauli spin blockade (PSB). These results open the way to perform hole spin qubit experiments in these devices.
\end{abstract}
\maketitle
\section*{Introduction}
\noindent Single hole spins confined in quantum dots in Ge/Si core/shell nanowires combine several advantageous properties which makes them potentially very powerful quantum bits~\cite{Loss98,Kloeffel13b}. The natural abundance of non-zero nuclear spins in both silicon and germanium is relatively small and can be further reduced to a negligible amount by isotopic purification. Furthermore, hole spins have no contact hyperfine interaction due to their p-type wavefunction. These properties make hole spin qubits in silicon and germanium resilient against dephasing via interaction with nuclear spins.

\noindent A particularly promising feature of hole spins in Ge/Si core/shell NWs is the nature of spin-orbit interaction (SOI) in this system. Confinement to one dimension gives rise to an effective SOI in the valence band, which is predicted to be both strong and tunable~\cite{Kloeffel11, Kloeffel17}, enabling fast all-electrical spin manipulation. An external electric field can be used to set the strength of this SOI. This promises the capability of electrical gating of the SOI, allowing to switch to a large SOI for high interaction strengths and fast quantum operations, or to turn off SOI for increased qubit coherence. Furthermore, this SOI results in a Land\'e g-factor that is locally tunable by external electric as well as magnetic fields~\cite{Maier13, Brauns16}. Local control over the g-factor makes it possible to selectively address individual spin qubits and allows for selective coupling to microwave cavities~\cite{Kloeffel13}.

\noindent Confining single holes in QDs enables to implement the basic ingredients of experimental quantum computation using spin qubits~\cite{Loss98}. Single QDs form the fundamental building blocks, and it is therefore imperative to be able to reliably form and characterize them~\cite{Brauns16a}. Moreover, a high level of control over the exact position and shape of individual QDs is required to accurately tune level splittings~\cite{Kloeffel11}, spin relaxation times~\cite{Hanson07, Camenzind17}, and tunnel coupling strengths.

\noindent In addition to single QDs, tunnel-coupled double QDs are of particular interest, since these are platforms for spin-to-charge conversion schemes facilitating spin read-out and coupling of spins to microwave cavities~\cite{Samkharadze18, Mi18, Landig17}. Spin states of double and triple QDs can be used as qubit encodings which are insensitive to fluctuations in environmental degrees of freedom~\cite{Taylor05, Taylor13}. Moreover, quantum operations on these qubits may be performed using different mechanisms than for single spin qubits, for instance only relying on the Heisenberg exchange interaction~\cite{Divincenzo00, Medford13}. Finally, double as well as triple QDs feature charge states with an increased dipole moment, potentially leading to enhanced coupling strengths of spin qubits to microwave cavities~\cite{Landig17}.\\

\noindent In this Letter, we demonstrate a large amount of control over the formation of single, double and triple QDs in Ge/Si NWs, all with a low hole occupation number. Using five bottom gate electrodes, we tune the size and position of single QDs defined in the NW. Furthermore, we form tunnel-coupled double and triple QDs. In the double QD configuration, we observe Pauli spin blockade~\cite{Ono02, Hanson07} (PSB).
\section*{Setup}
\noindent We use a Ge/Si NW~\cite{Conesa-Boj17} with an estimated Ge core radius of 10\,nm and Si shell thickness of 2.5\,nm (see Fig. 1a and b). Five Ti/Pd bottom gate electrodes are lithographically defined on a p++-doped Si substrate covered with 290\,nm thermal oxide. The bottom gates have a thickness of $\sim$15 nm, a width of 20\,nm, and are equally spaced with a pitch of 50\,nm. On either side of these gates, a plateau gate (green in Fig. 1b) is defined, which serves to prevent bending of the NW. The bottom gates are subsequently covered by a layer of $\text{Al}_2\text{O}_3$ of thickness 20\,nm through atomic layer deposition at $\SI{225}{\celsius}$. In a next step, the NW is placed deterministically on top of the bottom gates using a micromanipulator setup. Electrical contact to the NW is made through two Ti/Pd ($\sim$0.5/60\,nm) contact pads, which are lithographically defined and metallized after a brief HF dip to strip the NWs native oxide.\\

\noindent Due to the type-II staggered band alignment of silicon and germanium, a hole gas accumulates in the core~\cite{Lu05}. By applying positive voltages to the gate electrodes, the hole density can be depleted locally, resulting in the formation of quantum dots. We perform transport measurements by applying a dc source-drain bias $V_{SD}$ over the NW and measuring the differential conductance using standard lock-in techniques with a small ac excitation in the range of 20-100$\,\mu$V. All measurements were performed at a temperature of 1.4\,K, without application of an external magnetic field, and with the back gate grounded.
\section*{Single Quantum Dots}
\begin{figure}[t]
\includegraphics[width=0.48\textwidth]{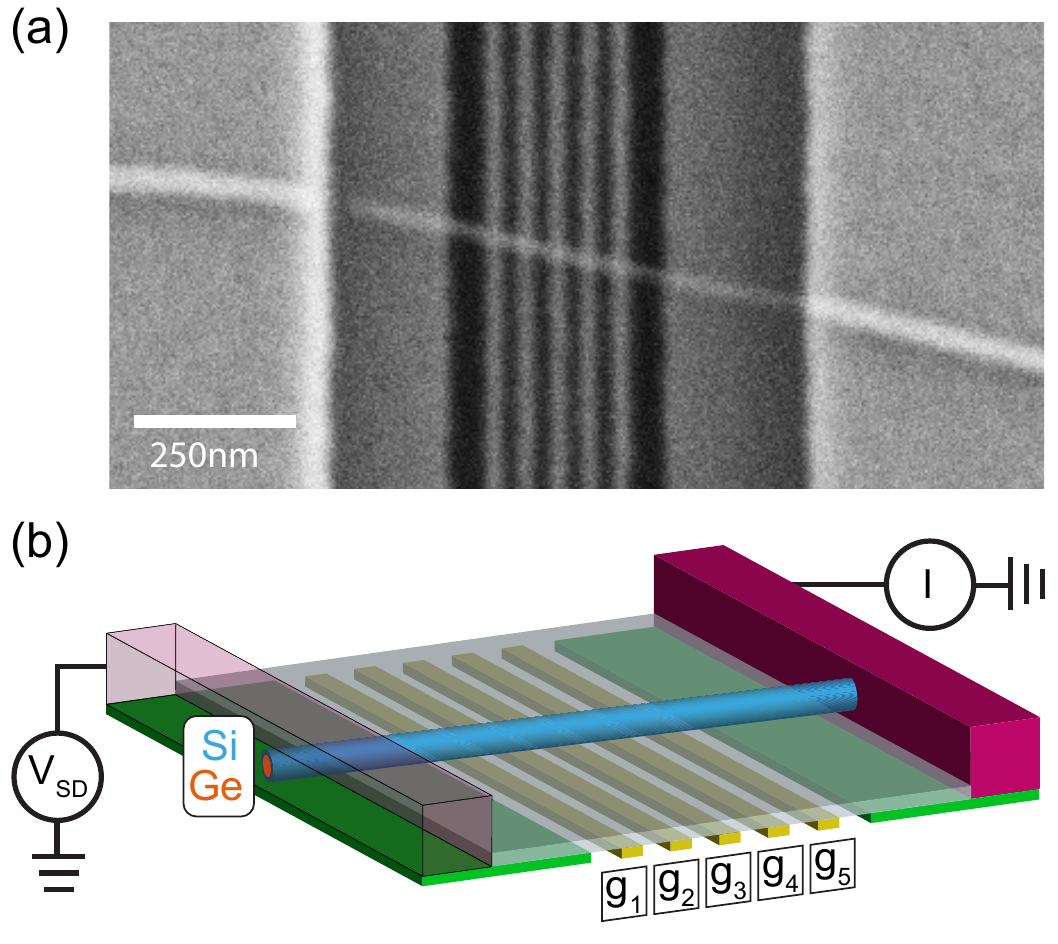}
\caption{(a). Scanning electron micrograph of a device similar to the one used in this work. (b) Schematic overview of device and measurement setup. The NW is shown in blue, with the core in orange, bottom gates are in yellow and green, and contacts in purple.}
\label{fig:Figure1}
\end{figure}
\noindent Figure 2a gives an overview of the different configurations of biased gates and dot sizes that were studied. QDs can be formed using two, three, four or five neighboring gates. For each dot size, the outer two gates (red in Fig. 2a) form tunnel barriers between the QD and the source and drain reservoirs. The voltage on individual or multiple middle gates (green in Fig. 2a) are used to tune the electrochemical potential of the QD. Unused gates (white in Fig. 2a) are grounded. In Figure 2b and c, examples of measured charge stability diagrams (Coulomb diamonds) are shown for the case of a single QD formed by two and three neighboring gates, respectively (see Fig. 2a, top panels). Similar Coulomb diamond measurements were made for larger QDs formed by four and five gates. In case of the QD defined by two adjacent gates, we find that sweeping the voltage on one of the two gates has a large effect on the tunnel barriers defining the dot. As a result, only a few charge transitions can be observed for this configuration. For the other dot sizes, the tunnel barriers are much less affected by the voltage on one of the middle gates, and we observe a large number of regular Coulomb diamonds.\\

\begin{figure}[t]
\includegraphics[width=0.48\textwidth]{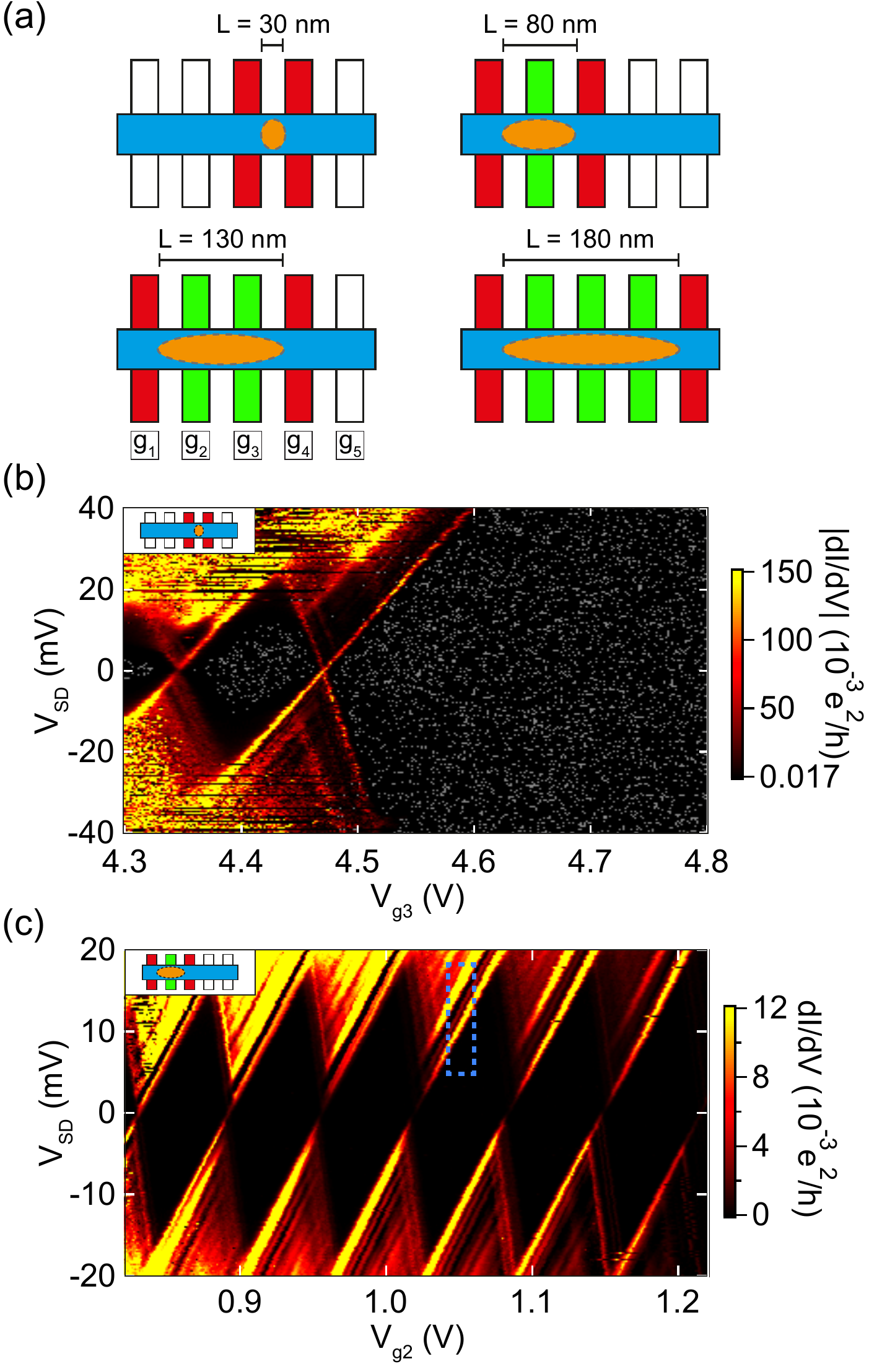}
\caption{(a). Schematic picture of the gate configurations used to form QDs (orange) of different lengths using 2, 3, 4 and 5 gates, respectively. (b) Lock-in signal $dI/dV$ versus $V_{SD}$ and $V_{g3}$ of QD formed by two gates. To enhance contrast, values below the colorscale were given a grey color. (c) Lock-in signal $dI/dV$ versus $V_{SD}$ and $V_{g2}$ of QD formed by three gates. Blue dashed rectangle shows an example of an averaging window used to extract excited state energies. Insets in (b) and (c) schematically show used gate configurations.}
\label{fig:Figure2}
\end{figure}
\noindent Table I summarizes parameters extracted from the Coulomb diamond measurements for the four different dot sizes. In Figure 3 values of the hole addition energy $E_{add}$ are plotted, which were extracted from the height of the Coulomb diamonds for the different dot sizes and for various hole occupation numbers. We find that $E_{add}$ is largest for the smallest dot and decreases for increasing dot size, in agreement with the expectation that both charging energy and orbital level splittings decrease with dot size. Going to larger hole occupation numbers also results in an overall decrease of $E_{add}$.\\

\noindent The conductance measurements feature additional resonances at higher values of $V_{SD}$. We extract energies for these resonances by averaging the difference of the first resonance and the ground state transition, in windows similar to the one drawn in Figure 2c. Here we convert the difference in $V_{SD}$ to energy using lever arms determined from the slopes of each Coulomb diamond. The third column of Table I lists typical energies $E_{orb}$ found in this way for the different dot sizes. Consistent with the level splitting of orbital hole states~\cite{Kouwenhoven01, Escott10}, $E_{orb}$ depends strongly on the longitudinal dot size, with smaller dots featuring higher values of $E_{orb}$. Note that incomplete knowledge of the exact confinement potential and the hole effective mass makes it difficult to compare our measurements to a theoretical model of orbital level splitting.\\

\noindent Furthermore, we estimate the lowest measurable hole occupation number $N_{est}$ for the different dot sizes by comparing the used gate voltages with pinch-off voltages obtained at high $V_{SD}$. For dots formed by 3 to 5 neighboring gates, we find relatively low occupation numbers ranging from 15 to 38 (see Table I). This method is not reliable for QDs defined by only two gates, since both gates directly define the tunnel barriers of the dot. However, there are several indications suggesting that the single hole occupation regime is reached in this case. First of all, the last Coulomb diamond edge visible in Figure 2b increases linearly up to at least $|V_{SD}|$ = 40\,mV. Furthermore, even at high $V_{SD}$, no features involving tunneling of multiple holes are observed for the last visible Coulomb diamond (which would appear as lines intersecting the diamond edges on the high gate voltage side). We do find multiple resonances in the last Coulomb diamond for low and high $V_{SD}$, which could arise from tunneling involving excited states. However, the splittings of these lines is lower than those found for the larger dots. Therefore, it is unlikely that these resonances correspond to tunneling involving excited orbital states in a small QD. Furthermore, we observe that the splitting of these resonances strongly depends on gate voltages applied to $g_2$ and $g_5$ (not shown), again making it unlikely that they correspond to excited orbital states~\cite{Mottonen10}. A likely explanation is that these lines arise from modulation of the reservoir density of states~\cite{Bjork04, Escott10, Mottonen10}. Finally, the energy of the first excited state in the second Coulomb diamond in Figure 2b (around $V_{g3}$\,=\,4.35\,V) appears to be significantly reduced with respect to that found in the last Coulomb diamond, consistent with an exchange energy appearing for two-hole states. More conclusive evidence of single hole occupation could be obtained by using a charge sensor~\cite{Hu07}.\\

\noindent Finally, we observed in multiple devices that QDs formed by three or more gates are vulnerable to splitting up when biasing the center gates too positively. This impedes reaching single-hole dot occupation for the larger dot sizes. Moreover, conductance becomes too low to measure when increasing the gate voltages, again potentially preventing the observation of single-hole occupation regimes. However, the conductance exceeds 0.1\,$e^2/h$ on the last diamond in Figure 2b, thus adding more evidence for the single-hole regime. 
\begin{table}[t]
\begin{tabular}
{|c|c|c|c|c|}
 \hline
 \# gates & $E_{add}$ (meV) & $E_{orb}$ (meV)& $L$ (nm) & $N_{est}$\\
 \hline
 \hline
 2 & 20 & 12.8 & 30 & 1, see main text\\
 3 & 17 & 4.8 & 80 & 15\\
 4 & 13 & 2.1 & 130 & 35\\
 5 & 10 & 1.3 & 180 & 38\\
 \hline
\end{tabular}
\caption{Typical extracted single dot parameters: addition energies $E_{add}$, excited state energies $E_{orb}$, lithographically defined distances $L$ between gates creating QD tunnel barriers, and estimated hole numbers $N_{est}$.}
\end{table}
\begin{figure}[t]
	\includegraphics[width=0.45\textwidth]{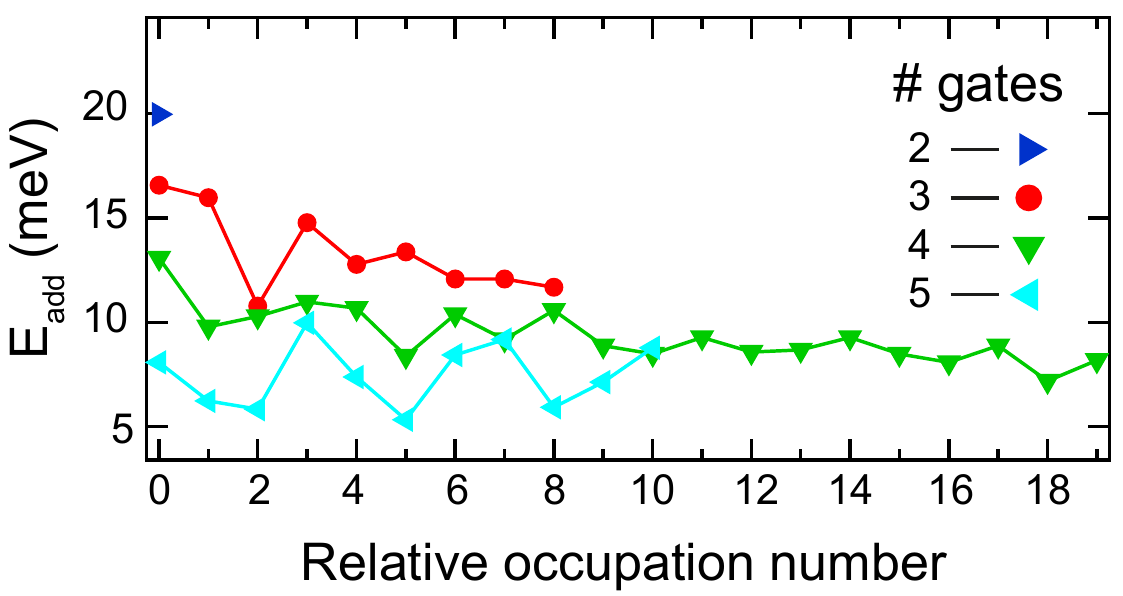}
	\caption{Extracted values of $E_{add}$ for various QD lengths as a function of relative occupation number.}
	\label{fig:Figure3}
\end{figure}
\section*{Double Quantum Dots}
\noindent Next, we demonstrate controllable formation of double QDs. As shown in the charge stability diagrams in Figure 4a, a single QD formed by five gates can be continuously split up into a double tunnel-coupled QD, by increasing the voltage on gate $g_3$. Here, the voltage on gates $g_2$ and $g_4$ are swept and the current through the NW is measured for each point. The leftmost charge stability diagram shows single-dot behavior, in which diagonal lines are Coulomb peaks corresponding to sequential addition of single holes to the dot. The middle panel shows a charge stability diagram of a double QD featuring high coupling between the dots, as evidenced by the bending of the charging lines. The right panel shows conductance only when the electrochemical potentials of the two dots are aligned, in the form of bias triangles~\cite{Vanderwiel02}. The absence of conductance along the charging lines indicates that significant cotunneling with the lead reservoirs can be avoided. These measurements indicate that we have a large amount of control over the capacitive coupling and tunnel-coupling between the two QDs. 
\subsection*{Pauli Spin Blockade}
\begin{figure*}[htb]
	\centering
	\includegraphics[width=0.95\textwidth]{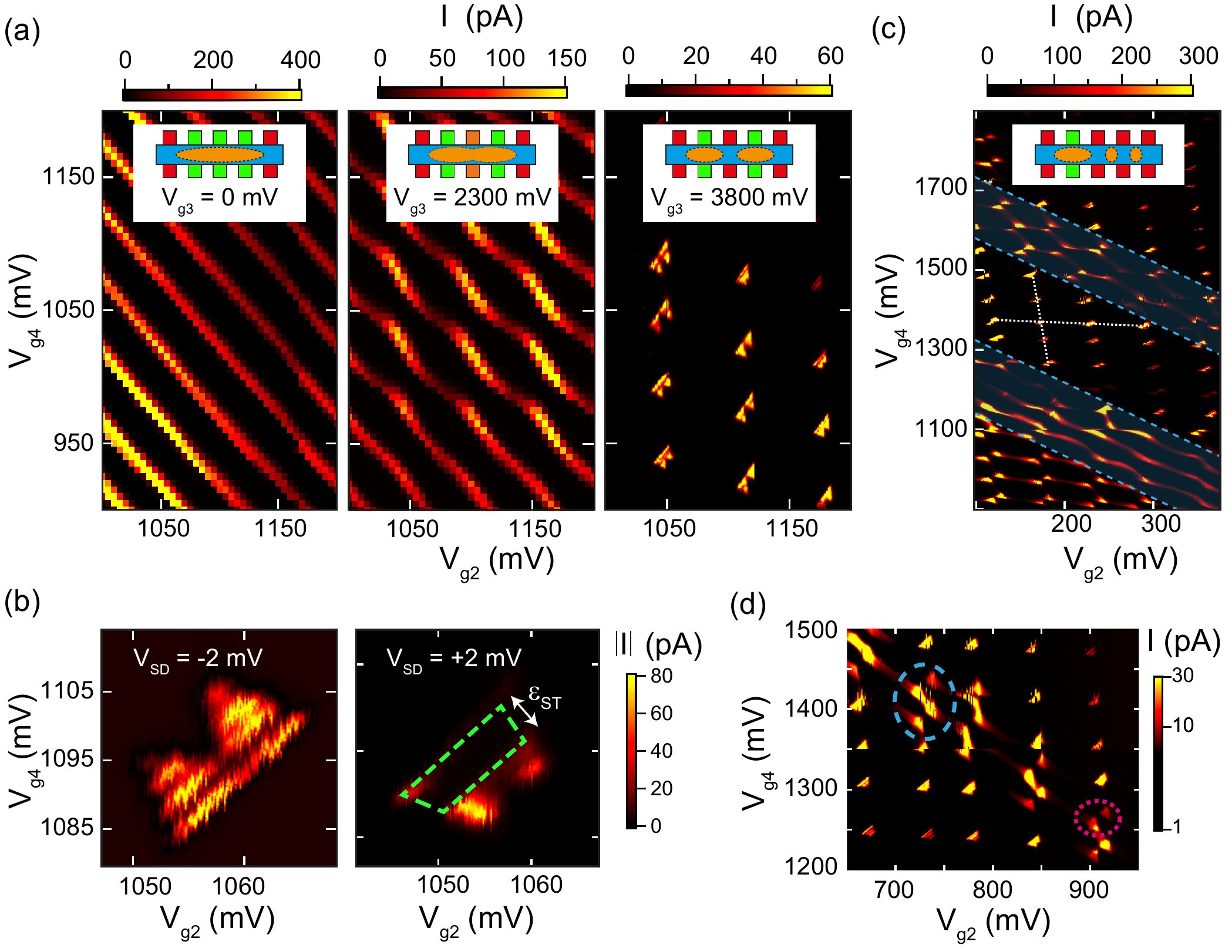}
	\caption{(a) Charge stability diagrams for different values of the voltage on $g_3$, showing a transition from a single QD to a double QD, at $V_{SD}$ = 2 mV. Insets schematically show QD configurations. (b) Zoom-in of a pair of bias triangles, at $V_{g3}$\,=\,3800\,mV. Plotted is the dc current for positive and negative $V_{SD}$. The strong reduction in the area enclosed by the dashed green line indicates the presence of Pauli spin blockade. (c) Charge stability diagram with highlighted (shaded blue regions) triple QD features. White dotted lines indicate the slope of charge transitions of the outer two dots. (d) Charge stability diagram of triple QD. Dashed blue and dotted pink circles highlight triple dot resonances. The jump in conductance around $V_{g4}$\,=\,1350\,mV originates from stitching two measurements together.}
	\label{fig:Figure4}
\end{figure*}
\noindent Pauli spin blockade is a basic ingredient of many spin qubit experiments, in which interdot transitions are blocked for spin triplet but not for singlet states~\cite{Ono02, Hanson07}. As such, it forms a means of reading out spin qubit states. When measuring the conductance through a double QD, the blockade may be observed for one sign of $V_{SD}$, but not for the other. In this work, the relevant spin states are those of Kramers doublets formed by mixed heavy hole and light hole states~\cite{Kloeffel11}.\\

\noindent Focussing on a single interdot transition, we observe Pauli spin blockade when measuring bias triangles for positive and negative $V_{SD}$ (see Fig. 4b). PSB shows up as a region of reduced conductance inside the bias triangles for one sign of $V_{SD}$ (indicated by dashed green line in Fig. 4b). Inside this region, current is suppressed by roughly a factor 10 for positive $V_{SD}$ compared to the case of negative $V_{SD}$. The size of the blockaded region is determined by the singlet-triplet splitting $\epsilon_{ST}$ in the single dots (see white arrow in Fig. 4b, right panel). We find $\varepsilon_{ST}$ to be 1\,meV, which compares well with other measurements~\cite{Brauns16b, Zarassi17}. Moreover, various processes may lift PSB, including spin-flip cotunneling, spin-flip reservoir exchange~\cite{Biesinger15}, hyperfine interaction, and SOI~\cite{Li15, Brauns16b, Zarassi17}. The resulting leakage current thus forms a probe to detect the strength of these processes. We observe leakage current that depends on the detuning of the electrochemical potentials in the two dots and on the magnitude of an applied magnetic field~\cite{Froning18}. Detailed study of leakage processes would go beyond the scope of this work, but can be performed in the future. 
\section*{Triple Quantum Dots}
\noindent We find that the double quantum dot can be further subdivided into a triple quantum dot, by increasing the voltage on $g_4$. In this case, the triple dot is likely composed of two small QDs between gate pairs $g_3$-$g_4$ and $g_4$-$g_5$, as well as a larger QD between $g_1$-$g_3$. In the charge stability diagram shown in Figure 4c, triple dot features show up as lines with enhanced conductance with an intermediate slope (see dashed blue lines). Figure 4d shows a zoomed-in region of the triple QD charge stability diagram in which additional features of enhanced conductance appear. Similar to bias triangles in a double QD, conductance is enhanced when the electrochemical potential of the center dot is aligned with that of one of the outer dots (dotted pink circle in Fig. 4d), or when the electrochemical potentials of all three dots are aligned (dashed blue circle in Fig. 4d)~\cite{Schroer07, Granger10}. The fact that we also observe conductance at points corresponding to DQD bias triangles suggests that there is cotunneling involving the center dot present in the measurements.
\section*{Conclusions and Outlook}
\noindent The demonstration of tunable single, double, and triple QDs opens the way to perform spin qubit experiments with few holes in these devices. Overall, we observe very good repeatability of the measurements, with gate voltage changes in the few-volt range having a negligible effect on device stability, as can be seen for instance in the large-range measurement of Figure 4c. These results enable several follow-up experiments. In particular, the strength and electric field dependence of the SOI could be determined from magnetic field dependence of leakage current in a double QD in the PSB regime~\cite{Li15, Zarassi17, Brauns16b, Froning18}. Moreover, we expect that a slightly different gate design than that presented here will enable reaching single-hole occupation in a controllable way. The use of a charge sensor defined in vicinity of the QD under study would aid in this endeavour. 
\section*{Acknowledgements}
\noindent We thank C. Kloeffel, D. Loss, and M. Ran\v ci\'c for helpful discussions. We acknowledge the support of the Swiss National Science Foundation (Ambizione Grant No. PZOOP2161284/1) and Project Grant No. 157213), the Swiss Nanoscience Institute, the European Microkelvin Platform EMP, the NCCR Quantum Science and Technology (QSIT), and the Netherlands Organization for Scientific Research (NWO).

\end{document}